\documentclass[journal]{IEEEtran}
\usepackage{amsmath}
\usepackage{amsfonts}
\usepackage{graphicx}

\ifCLASSOPTIONcompsoc \usepackage[caption=false,font=normalsize,labelfon
t=sf,textfont=sf]{subfig}
\else
\usepackage[caption=false,font=footnotesize]{subfig}
\fi

\begin{document}
\pagenumbering{gobble}

\title{Non-unique water and contrast agent solutions in dual-energy CT}
\author{JP Phillips, Emil Y. Sidky, Fatma Terzioglu, 
        Ingrid S. Reiser, Guillaume Bal, Xiaochuan Pan

        \thanks{J. P. Phillips, E.Y Sidky, I. S. Reiser, and X. Pan are with the Department of Radiology MC-2026, The University of Chicago, 5841. S. Maryland Ave., Chicago, IL, 60637.}
        \thanks{F. Terzioglu is with the Department of Mathematics, North Carolina State University, 2311 Stinson Dr., Raleigh, NC, 27695}
        \thanks{G. Bal is with the Departments of Statistics and Mathematics, The University of Chicago, 5747 S. Ellis Ave., Chicago, IL, 60637}
        }
\maketitle

\begin{abstract}
The goal of this work is to study occurrences of non-unique solutions in dual-energy CT (DECT) for objects containing water and a contrast agent. Previous studies of the Jacobian of nonlinear systems identified that a vanishing Jacobian determinant indicates the existence of multiple solutions to the system. Vanishing Jacobian determinants are identified for DECT setups by simulating intensity data for practical thickness ranges of water and contrast agent. Once existence is identified, non-unique solutions are found by simulating scan data and finding intensity contours with that intersect multiple times. With this process non-unique solutions are found for DECT setups scanning iodine and gadolinium, including setups using tube potentials in practical ranges. Non-unique solutions demonstrate a large range of differences and can result in significant discrepancies between recovered and true material mapping.

\end{abstract}

\IEEEpeerreviewmaketitle

\section{Introduction}
\IEEEPARstart{D}{ual}-energy CT (DECT) systems are able to identify two different materials in a scanned object by simultaneously collecting two different spectral measurements. Hounsfield introduced the concept of DECT in the 1970s with the mathematical framework being developed in 1976 by Alvarez and Macovski \cite{hounsfield}\cite{alvarez_energy-selective_1976}. Their approach for material decomposition depends on the energy dependence of a material's linear attenuation coefficients, a property that can be exploited by making measurements with two different x-ray spectra.

This work focuses on projection-based material decomposition in DECT which requires solving a nonlinear system of equations for x-ray paths. When determining  the value of projection line integrals, the nonlinearity of the mapping from integral to DECT measurements results in the possibility of non-unique line integral values. Levine first demonstrated an example of non-unique solutions for a water and bone material basis using a spectra of three discrete photon values \cite{levine_nonuniqueness_2017}. Alvarez later investigated uniqueness in relation to a nonvanishing Jacobian and found in general that only local uniqueness is guaranteed by a nonvanishing Jacobian \cite{alvarez_invertibility_2019}. This work seeks to demonstrate non-uniqueness when using basis materials of water and CT contrast agents iodine and gadolinium. In addition to using different basis materials from past studies, this work will also look for non-unique solutions using full x-ray spectra rather than discrete photon energies.

 \section{Methods}
For contrast imaging with a DECT system, we assume that only water and the contrast agent are present in the scanned object. Such an object is scanned using x-ray spectra $S_1(E)$ and $S_2(E)$ which correspond to a low and high x-ray tube potentials respectively. Using the publicly available Python package SpekPy, x-ray spectra can be modeled for a given tube potential \cite{spekpy}. For such sources, the total number of photons transmitted through an object is

\begin{equation}
    I_i(t_w, t_c) = \int_{0}^{E_{max}} S_i(E) e^{-\mu _w(E)t_w - \mu _c(E)t_c} dE, \; i = 1,2
\end{equation}

\noindent where $\mu_w(E)$ and $\mu_c(E)$ represent the energy dependant attenuation coefficients of water and the contrast agent respectively while $t_{w}$ and $t_{c}$ are the thickness of water and contrast agent respectively in centimeters. X-ray spectra are normalized so that intensity is measured relatively and will only have values between 0 and 1.

The nonlinearity of the system in Eq. 1 gives rise to non-uniqueness, resulting in different pairs of $t_w$ and $t_c$ values producing the same intensity values. Non-uniqueness can be prevented if parameters are selected that ensure the measured intensity values can only be obtained with one set of thicknesses. The Jacobian of the system can be used to guide parameter selection.

For notation, we will define

\begin{equation*}
    \mu = \begin{bmatrix}
            \mu_w(E)\\
            \mu_c(E) 
        \end{bmatrix}, \;\;
    t = \begin{bmatrix}
            t_w\\
            t_c
        \end{bmatrix}
\end{equation*}

When looking at the Jacobian, we will consider the negative logarithm of the intensity measurements defined as

\begin{equation}
    g_i(t) = - \ln I_i(t), \; i = 1,2
\end{equation}

For the system in Eq. 1, the entries of Jacobian matrix $J(t)$ are given by

\begin{equation}
    J_{ij}(t) = \frac{\partial g_i}{\partial t_j}(t) = \frac{\int_{0}^{E_{max}} S_i(E) \mu_j(E)e^{-\mu \cdot t} dE}{\int_{0}^{E_{max}} S_i(E)e^{-\mu \cdot t} dE}, \; i,j = 1,2
\end{equation}

The matrix from Eq. 3 is then used to calculate the Jacobian determinant
\begin{equation}
    det(J) = J_{11}(t)J_{22}(t) - J_{12}(t)J_{21}(t)
\end{equation}

\begin{figure*}[t!]
\centering
\subfloat{\includegraphics[width=3in]{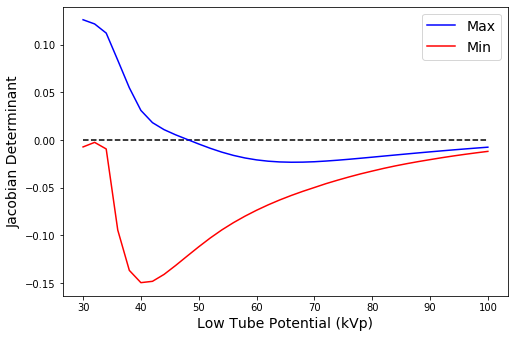}}
\hfil
\subfloat{\includegraphics[width=3in]{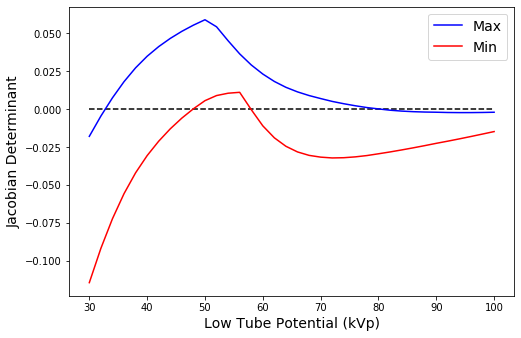}}
\caption{Plots of Jacobian determinant for thickness sweep. The blue and red curves correspond to the maximum and minimum values of the determinant among all thickness combinations. High tube potential was held at 120 kVp and low tube potential was varied from 30 to 100 kVp. Materials were water and iodine (left) or gadolinium (right).}
\end{figure*}

It has previously been proved that should Eq. 4 not equal zero for all $t \in \mathbb{R}$, then the estimation of $t$ from $I(t)$ is unique \cite{gale_jacobian_1965}\cite{bal_uniqueness_2020}. For this work we demonstrate the dependence of the Jacobian determinant on material basis pair and x-ray spectra. For high tube potentials of 120 kVp and 140 kVp, values of $I_i(t_w, t_c)$ were simulated for $t_w$ varied between 0 and 30 cm and $t_c$ varied between 0 and 3 cm for iodine and gadolinium with a 1\% concentration. Low tube potential was varied from 30 to 100 kVp for the simulation. Eq. 3 and Eq. 4 were used to calculate the Jacobian determinant for every pair of intensities measured for a low and high tube potential combination. If zero is contained between the minimum and maximum values of the Jacobian determinants for a tube potential pair, then there exists a non-unique solution.

To search for the values of non-unique solutions, a grid search of solutions to Eq. 1 for values of $I_i(t_w, t_c)$ ranging from 0 to 1 was completed. For a given pair of $I_i(t_w, t_c)$ values, contour curves specified by $t_w$ and $t_c$ were drawn. $t_w$ and $t_c$ were restricted to ranges of 0 to 30 cm and 0 to 3 cm respectively. If the contours of $I_1$ and $I_2$ intersected once, then there was a set of thicknesses that resulted in measuring the specified intensity values. If the contours intersected twice, than there were two pairs of thicknesses that would solve Eq. 1 for the values of $I_1$ and $I_2$ and the solutions would not be unique.

From the study of the Jacobian determinant, we know that changing the x-ray spectra and materials will effect the existence of non-unique solutions. The results of examining the Jacobian determinant are useful for parameter selection as they highlight where non-unique solutions are likely to appear. Based on the Jacobian, for iodine the low tube potential was varied from 30 to 50 kVp and the high tube potential from 70 to 140 kVp. For gadolinium, the low tube potential was varied from 40 to 80 kVp and the high tube potential from 100 to 140 kVp.

\section{Results}

\begin{figure}[t!]
    \centering
    \includegraphics[width=8cm]{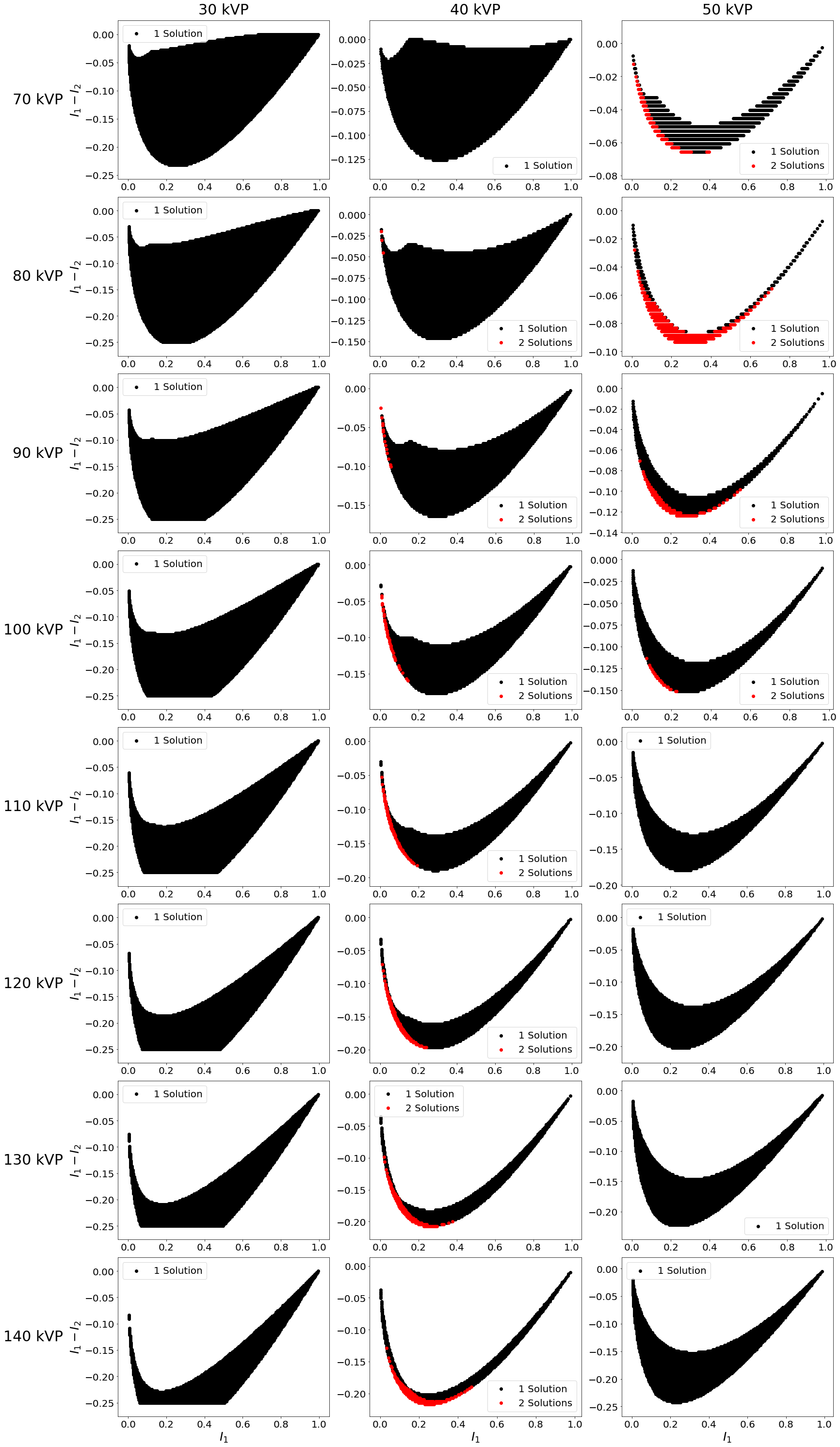}
    \caption{Intensity sweep plots for water and iodine. Low tube potential increases from 30 kVp to 50 kVp moving left to right and high tube potential increases from 70 kVp to 140 kVp moving top to bottom. Black areas indicate intensity combinations that only have one solution to Eq. 1 and red areas indicate where there are two solutions.}
\end{figure}

\begin{figure*}[t!]
    \centering
    \includegraphics[width=17cm]{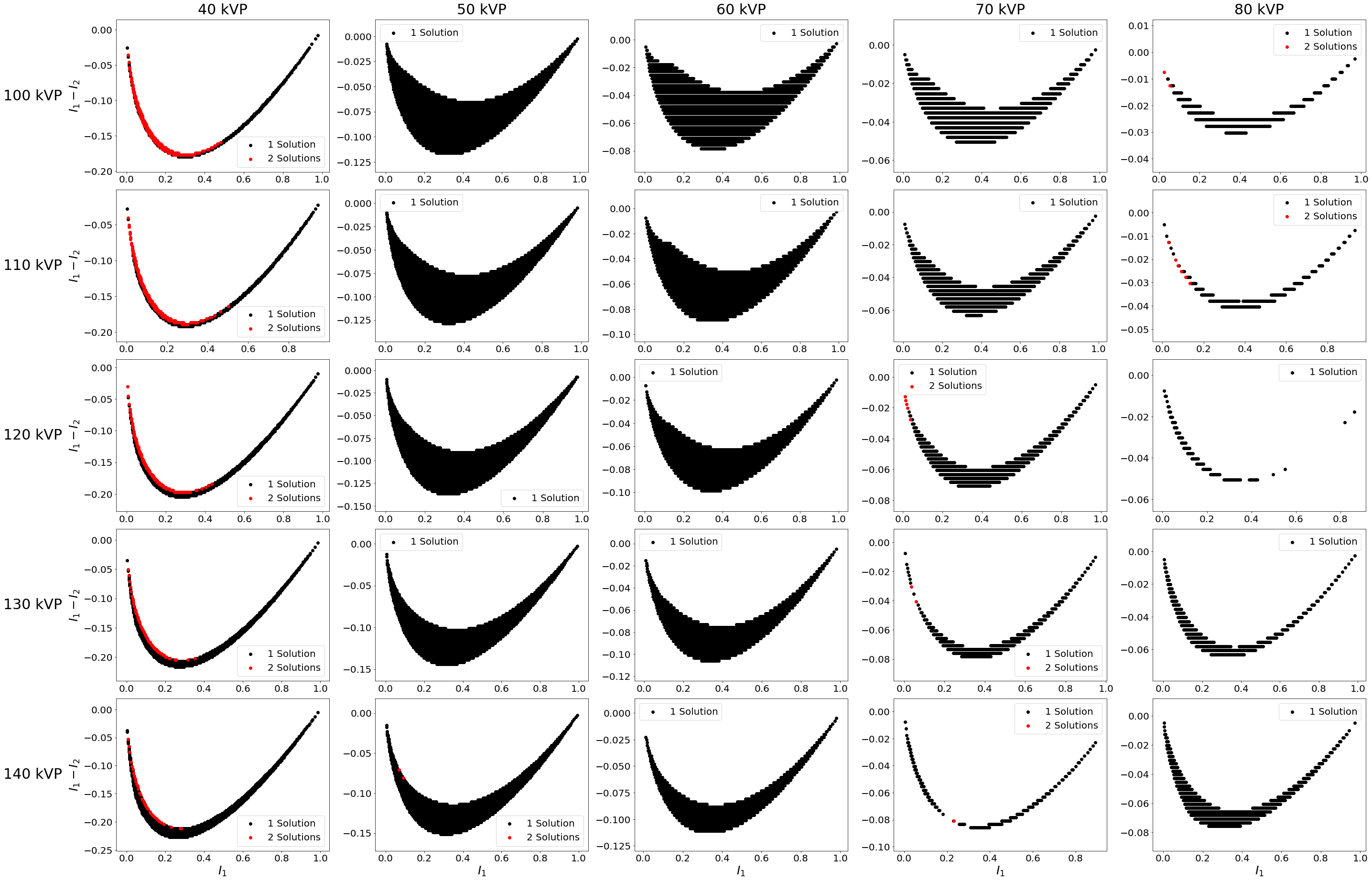}
    \caption{Intensity sweep plots for water and gadolinium. Low tube potential increases from 40 kVp to 80 kVp moving left to right and high tube potential increases from 100 kVp to 140 kVp moving top to bottom. Black areas indicate intensity combinations that only have one solution to Eq. 1 and red areas indicate where there are two solutions.}
\end{figure*}

\begin{figure*}[t!]
    \centering
    \includegraphics[width=18cm]{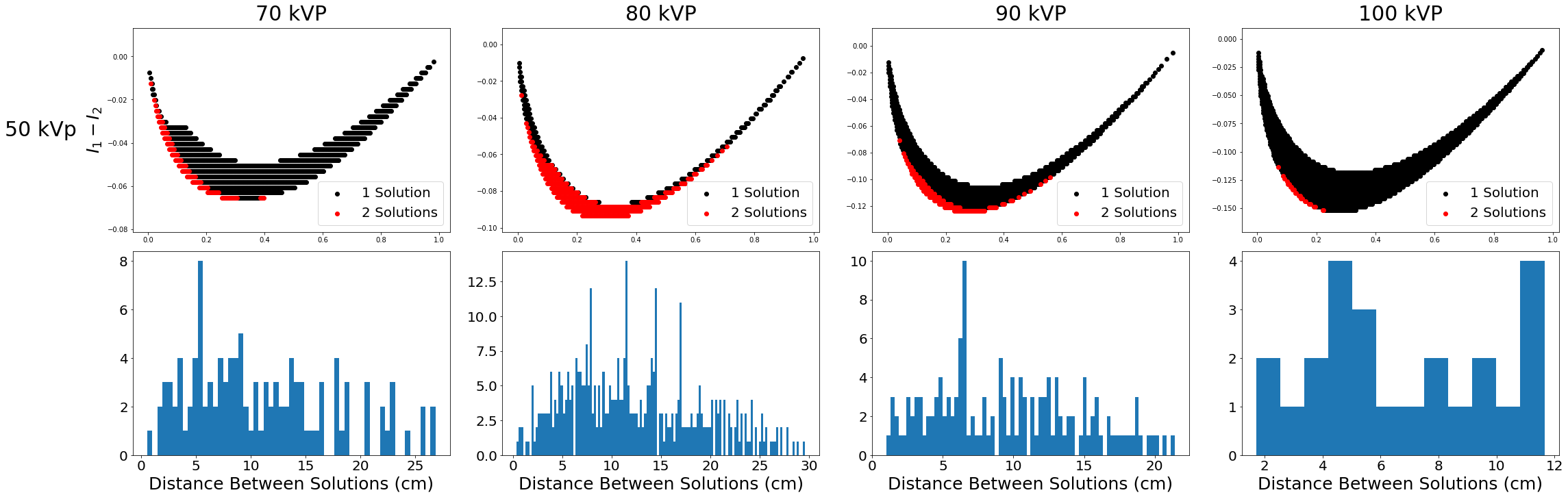}
    \caption{Selected solution plots for water and iodine. Low tube potential is held at 50 kVp and high tube potential increases from 70 to 100 kVp from left to right. The same solution plots from Fig. 2 are shown (top) along with corresponding histograms of the Euclidean distance between each pair of solutions for a given intensity pair (bottom).}
\end{figure*}

Using Eq. 3 and Eq. 4, values of the Jacobian determinant were calculated. Figure 1 shows how the Jacobian determinant changes for water-iodine and water-gadolinium material pairs as low tube potential is varied while high tube potential is held constant at 120 kVp. Since non-unique solutions occur when the Jacobian determinant vanishes, the plots in Figure 1 suggest that multiple solutions are likely to be found for low tube potentials in the 30-50 kVp range for iodine and between the 35-50 kVp and 60-80 kVp ranges for gadolinium.

\begin{figure*}[t!]
    \centering
    \includegraphics[width=18cm]{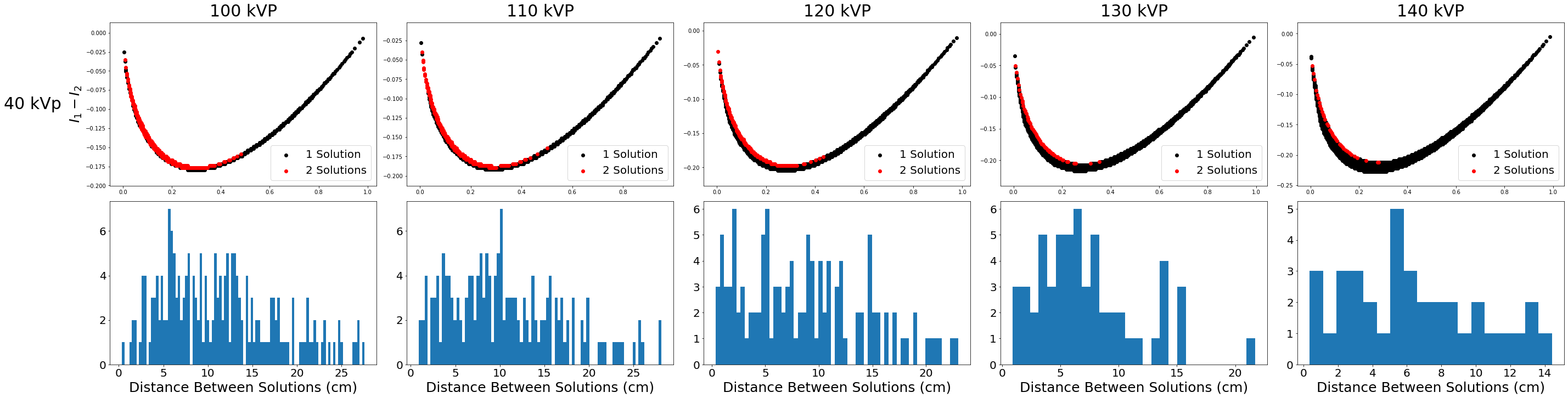}
    \caption{Selected solution plots for water and gadolinium. Low tube potential is held at 40 kVp and high tube potential increases from 100 to 140 kVp from left to right. The same solution plots from Fig. 3 are shown (top) along with corresponding histograms of the Euclidean distance between each pair of solutions for a given intensity pair (bottom).}
\end{figure*}

All possible solutions to Eq. 1 for given sets of tube potentials when measuring 0 to 30 cm of water and 0 to 3 cm of iodine are shown in Figure 2. Similar results for gadolinium are displayed in Figure 3. For both figures sparsity of solutions in some plots is due to discrete sampling of intensities. Multiple tube combinations resulted in non-unique solutions as denoted by red regions on the plots. For iodine non-uniqueness occurred for a low tube potential of 40 kVp and high tube potential between 100 and 140 kVp. Non-uniqueness also occurred for a low tube potential of 50 kVp and high tube potential between 70 and 100 kVp. When using gadolinium as the contrast agent, non-uniqueness occurred when using a low tube potential of 40 kVp and all high tube potentials studied. For low tube potentials of 70 and 80 kVp, a small number of non-unique solutions were found. While small in number, these points are significant as they occur in the practical range of DECT which typically use 70 to 100 kVp and 135 to 150 kVp \cite{tatsugami_dual-energy_2022}.

The red points in Figure 2 and Figure 3 can be plotted on a plane defined by $t_w$ and $t_c$. The Euclidean distance between the points $d$ can then be calculated with

\begin{equation}
    d = \sqrt{(t_w^{I_1} - t_w^{I_2})^2 + (t_c^{I_1} - t_c^{I_2})^2}
\end{equation}

The farther away the points are, the greater the difference between the values of the solutions. Figure 4 displays the solutions plots for iodine using a 50 kVp low tube potential and corresponding histograms showing $d$ for all two solution points. Figure 5 shows similar results for gadolinium using a 40 kVp low tube potential. The histograms for iodine and gadolinium indicate that some intensities will result in two solutions that are significantly different.

\section{Conclusion}
The main results from this work identify DECT parameters that can result in non-unique material mapping when scanning an object containing water and iodine or gadolinium. Most significant is the identification of multiple solutions when using practical tube potentials that are used in a clinical setting. Furthermore, the discrepancy between possible solutions can be large and result in a significant error in material mapping.

This work will be extended to examine non-uniqueness in the practical range more closely and look for non-uniqueness for more material combinations. Results using a finer sampling of some parameter combinations will also be presented at the conference.

\section*{Acknowledgment}
This research was supported in part by the National Institute
of Biomedical Imaging and Bioengineering of the National
Institutes of Health under award numbers R01EB023968 and
R21CA263660.

\ifCLASSOPTIONcaptionsoff
  \newpage
\fi

\bibliography{./IEEEexample}

\end{document}